\documentclass{article}

\usepackage{cite}
\usepackage{amsthm,amsmath,amssymb,amsfonts}
\usepackage{algorithm}
\usepackage[noend]{algorithmic}
\usepackage{graphicx}
\usepackage{textcomp}
\usepackage{xcolor}
\usepackage{url}
\usepackage{tikz}
\usepackage{verbatim}
\usepackage{hyperref}

\usepackage{pgfplots}
\pgfplotsset{compat = 1.18}

\def\BibTeX{{\rm B\kern-.05em{\sc i\kern-.025em b}\kern-.08em
    T\kern-.1667em\lower.7ex\hbox{E}\kern-.125emX}}

\newtheoremstyle{mystyle}
  {}                                            
  {}                                            
  {}                                            
  {}                                            
  {\bfseries}                                   
  {}                                            
  { }                                           
  {\thmname{#1}\thmnumber{ #2}\thmnote{ (#3)}.} 

\def\verbatimboxed#1{\begingroup

\setbox0=\vbox{\parskip=0pt\topsep=0pt\partopsep=0pt\hsize=0.9\linewidth
\verbatiminput{#1}}%
\begin{center}\fbox{\box0}\end{center}%
\endgroup}
  
\theoremstyle{mystyle}
\newtheorem{definition}{Definition}

\begin{document}

\newcommand{\toolname}{Gamu Blue}


\title{\toolname{}: A Practical Tool for Game Theory Security Equilibria \footnote{This paper has been accepted for publication by IEEE under copyright © 2024 IEEE.}}

\author{
    Ameer Taweel\\
    \texttt{Koç University}
    \and
    Burcu Yıldız\\
    \texttt{EPFL}
    \and
    Alptekin Küpçü\\
    \texttt{Koç University}
}

\maketitle


\begin{abstract}
    The application of game theory in cybersecurity enables strategic analysis, adversarial modeling, and optimal decision-making to address security threats' complex and dynamic nature. Previous studies by Abraham et al. and Biçer et al. 
    presented various definitions of equilibria to examine the security aspects of games involving multiple parties. Nonetheless, these definitions lack practical and easy-to-use implementations. Our primary contribution is addressing this gap by developing \toolname{}, an easy-to-use tool with implementations for computing the equilibria definitions including $k$-resiliency, $\ell$-repellence, $t$-immunity, $(\ell,t)$-resistance, and $m$-stability. 
\end{abstract}


\section{Introduction}


Game theory provides a powerful toolkit for understanding, analyzing, and addressing the complex challenges in cybersecurity. Cybersecurity researchers and professionals use game theory for strategic analysis, adversarial modeling, optimal decision-making, risk/vulnerability analysis, collaborative security, and protocol/mechanism design \cite{distributed-systems,ioc-game,network-security, distributed-file-sharing, vehicular-networks}. By applying game-theoretic concepts, one can analyze the interactions between attackers and defenders, identify optimal strategies, and understand the implications of different decisions in a dynamic and adversarial environment. 
It also facilitates collaborative approaches where multiple parties can coordinate their actions to enhance security. Moreover, game-theoretic principles guide the design of secure protocols and mechanisms that incentivize desired behaviors and discourage malicious activities \cite{new-definitions,ioc-game, network-security, distributed-file-sharing,incentivizing-outsourced-computation,ioc-smartcontract}.


Security solutions based on game theory rely on the rational behavior of the parties involved. In scenarios involving multiple parties, it is often safe to assume that individuals will not engage in malicious actions if those actions are detrimental to themselves. As a result, one can focus on addressing threats that are advantageous to the attacker rather than trying to countermeasure against all possible threats.


Previously, \cite{old-definitions} and \cite{new-definitions} presented equilibria definitions for analyzing games' security in the multi-party setting, but there are not any existing implementations to compute these equilibria. Our main contribution is developing \toolname{}, an easy-to-use tool with implementations for computing these equilibria. Our tool allows game theory and cybersecurity researchers to analyze their games using the security-related game-theoretic equilibria definitions. Moreover, it provides a baseline to compare future algorithmic improvements against previous ones.


\section{Preliminaries}

\subsection{Basics}

\begin{definition}[$n$-Player Normal-Form Game]
    An $n$-player perfect information normal-form game can be defined by a set of players $N = \{ p_1, \dots, p_n \}$, a set of actions $A_i = \{ a_i^1, \dots, a_i^{m_i} \}$ for each player $i$, and a utility function $U_i$ for each player $i$ that gives the utility of player $i$, given the actions chosen by all players. All players have complete and perfect knowledge of the game's state, including the other players' utility functions and available actions.

    Strategy $s_i$ of player $i$ is a probability distribution over the possible actions of the player $i$. $s_i$ is a pure strategy if it always chooses the same action. Otherwise, $s_i$ is a mixed strategy. We will only consider pure strategies in this paper because our algorithms rely on exhaustive search, and there are infinitely many mixed strategies.
    $s_i(a)$ is the probability the strategy $s_i$ assigns to the action $a$. $s_D = \{s_{p_a}, \dots, s_{p_b}\}$ is a strategy profile (an ordered set) containing the strategies of all players in the set $D = \{p_a, \dots, p_b\}$. $s_{-i}$ is the strategy profile of all the players excluding player $i$. $s_{-D}$ the strategy profile of all the players that are not in $D$. Given a strategy profile $s_N$ , we denote by $U_i(s_N)$ the expected utility of the player $i$, if the players in the game play the strategy profile $s_N$.
\end{definition}

\begin{definition}[Nash Equilibrium]
    For any game with the player set $N$, a strategy profile $s_N = \{s_{p_1}, \dots s_{p_n}\}$ is a Nash equilibrium if $\forall i \in N \ \forall s_i^\prime \neq s_i \ U_i(s_N) \geq U_i(s_i^\prime \cup s_{-i})$.

    Nash equilibrium is a fundamental concept in game theory that describes a state in which no player has an incentive to unilaterally deviate from their chosen strategy, given the strategies of other players.
    John Nash proved in \cite{nash-equilibrium-existence} that every finite game (games with a finite number of players and actions per player) has at least one Nash equilibrium.
\end{definition}

\begin{definition}[Weakly Dominant Strategy of a Player]
    A strategy $s_i$ of a player $i$ is its weakly dominant strategy if for all strategies $s_i^\prime \neq s_i$ of $i$ and for all strategy profiles $s_{-i}^\prime$ of players other than $i$, $U_i(s_i \cup s_{-i}^\prime) \geq U_i(s_i^\prime \cup s_{-i}^\prime)$. A weakly dominant strategy is not guaranteed to exist.
\end{definition}

\begin{definition}[Coalition Utility]
    Let $s_N$ be a strategy profile for all players in a game. Then, the utility $U_C(s_N)$ of a coalition $C$ is defined as $U_C(s_N) = \sum_{i \in C} U_i(s_N)$.

    This coalition utility definition simplifies the analysis by abstracting out how participants share utility internally.
\end{definition}

\begin{definition}[Weakly Dominant Strategy of a Coalition]
    A strategy profile $s_C$ of a coalition $C$ is its weakly dominant strategy if for all strategy profiles $s_C^\prime \neq s_C$ of $C$ and for all strategy profiles $s_{-C}^\prime$ of the players outside $C$: \\ $U_C(s_C \cup s_{-C}^\prime) \geq U_C(s_C^\prime \cup s_{-C}^\prime)$.
\end{definition}

\subsection{Multi-Party Secure Equilibria Definitions Introduced In \cite{old-definitions}}

\begin{definition}[$k$-resiliency]
    Given a player set $C \subseteq N$,  $s_C$ is a group best response for $C$ to $s_{-C}$, if for all strategies $s_C^\prime$ played by $C$ and $\forall i \in C$, we have $U_i(s_C \cup s_{-C}) \geq U_i(s_C^\prime \cup s_{-C})$. A joint strategy $s_N$ is a $k$-resilient equilibrium, if $\forall C \subseteq N$ with $|C| \leq k$, $s_C$ is a group best response for $C$ to $s_{-C}$, where $s_N = s_C \cup s_{-C}$.

    The authors of \cite{old-definitions} proposed $k$-resiliency to protect the mechanism outcome against a single coalition. If $s_N$ is $k$-resilient, it is secure against any single coalition of size less than or equal to $k$ because no such coalition can increase the utility of any of its members by deviating from $s_N$.
\end{definition}

\begin{definition}[$t$-immunity]
    A joint strategy $s_N$ is a $t$-immune equilibrium, if $\forall T \subseteq N$ with $|T| \leq t$, for all strategies $s_T^\prime$ played by the players in $T$, and $\forall i \notin T$, we have $U_i(s_T^\prime \cup s_{-T}) \geq U_i(s_N)$, where $s_N = s_T \cup s_{-T}$.

    \cite{old-definitions} proposed $t$-immunity to protect the mechanism outcome against Byzantine players. If $s_N$ is $t$-immune, it is secure against up to $t$ Byzantine players because such players cannot decrease the utility of any other player by deviating from $s_N$.
\end{definition}

\begin{definition}[$(k, t)$-robustness]
    A joint strategy $s_N$ is a $(k, t)$-robust equilibrium, if $\forall C \subseteq N$, $\forall T \subseteq N$ such that $C \cap T = \emptyset$, $|C| \leq k$, and $|T| \leq t$, for all strategies $s_T^\prime$ played by the players in $T$, and for all strategies $s_C^\prime$ played by the players in $C$, and $\forall i \in C$, we have $U_i(s_{-T} \cup s_T^\prime) \geq U_i(s_{-(C \cup T)} \cup s_C^\prime \cup s_T^\prime)$, where $s_N = s_C \cup s_T \cup s_{-(C \cup T)}$.

    The authors of \cite{old-definitions} proposed $(k, t)$-robustness to protect the mechanism outcome against a single coalition and Byzantine players. If $s_N$ is $(k, t)$-robust, it is secure against up to $t$ Byzantine players and any single rational coalition of size up to $k$: the Byzantine players cannot harm the other players by deviating from $s_N$, and no single coalition can increase the utility of any of its members by deviating from $s_N$.
\end{definition}

\subsection{Improved Multi-Party Secure Equilibria Definitions Introduced In \cite{new-definitions}}

\begin{definition}[$\ell$-repellence]
    Given a player set $C \subseteq N$, $s_C$ is a best collective response for $C$ to $s_{-C}$, if for all strategies $s_C^\prime$ played by $C$, we have $U_C(s_C \cup s_{-C}) \geq U_C(s_C^\prime \cup s_{-C})$. A joint strategy $s_N$ is an $\ell$-repellent equilibrium, if $\forall C \subseteq N$ with $|C| \leq \ell$, $s_C$ is a best collective response for $C$ to $s_{-C}$, where $s_N = s_C \cup s_{-C}$.

    The authors of \cite{new-definitions} proposed $\ell$-repellence to replace $k$-resiliency. $\ell$-repellence analyzes the security of a mechanism against the presence of a single coalition. If $s_N$ is $\ell$-repellent, it is secure against any single coalition of size up to $\ell$ because no such coalition can increase its \textit{total utility} by deviating from $s_N$. Note that $\ell$-repellence improves $k$-resiliency because it employs the transferable utility assumption \cite{transferable-utility} and abstracts out how the utility is shared among the participants internally. Thus, rather than considering the individual members of a coalition, it considers the total benefit of the coalition.
\end{definition}

\begin{definition}[$(\ell, t)$-resistance]
    A joint strategy $s_N$ is an $(\ell, t)$-resistant equilibrium, if $\forall C, T \subseteq N$ s.t. $C \cap T = \emptyset$, $|C| \leq \ell$, and $|T| \leq t$, for all strategies $s_T^\prime$ played by the players in $T$, and for all strategies $s_C^\prime$ played by $C$, we have $U_C(s_{-T} \cup s_T^\prime) \geq U_C(s_{-(C \cup T)} \cup s_T^\prime \cup s_T^\prime)$, where $s_N = s_C \cup s_T \cup s_{-(C \cup T)}$.

    \cite{new-definitions} proposed $(\ell, t)$-resistance to replace $(k, t)$-robustness. $(\ell, t)$-resistance analyzes the security of a mechanism against the presence of a single coalition and Byzantine players. If $s_N$ is $(\ell, t)$-resistant, it is secure against up to $t$ Byzantine players and any single coalition of size up to $\ell$: the Byzantine players cannot harm the other players by deviating from $s_N$, and no single coalition can increase its total utility by deviating from $s_N$. $(\ell, t)$-resistance improves over $(k, t)$-robustness by employing the transferable utility assumption.
\end{definition}

\begin{definition}[$m$-stability]
    A joint strategy $s_N$ is an $m$-stable equilibrium, if for all natural numbers $p \leq |N|$ and for all coalitions $C_1, \dots, C_p$ satisfying the following conditions:
    \begin{enumerate}
        \item $1 \leq |C_1|, \dots, |C_p| \leq m$
        \item $C_1 \cup \dots \cup C_p = N$
        \item $\forall i, j \in \{1, \dots, p\}$ such that $i \neq j$, $C_i \cap C_j = \emptyset$
    \end{enumerate}
    we have that for each $i = 1, \dots, p$, the strategy $s_{C_i}$ is weakly dominant for the coalition $C_i$ and that $s_{C_1} \cup \dots \cup s_{C_p} = s_N$.

    The authors of \cite{new-definitions} proposed $m$-stability to analyze the security of a mechanism against the presence of multiple coalitions. If $s_N$ is $m$-stable, it is secure against any number of coalitions of size up to $m$ because no coalition has incentive to deviate from $s_N$, even if it expects others to deviate from $s_N$. Like other definitions introduced in \cite{new-definitions}, $m$-stability employs the transferable utility assumption. \cite{new-definitions} proved that $(n - 1)$-stability implies $(n - 1)$-repellence and $(\ell, n - \ell)$-resistance for all $\ell \leq n - 1$.
\end{definition}


\section{Algorithms}

We mainly focus on the optimization problem with one strategy profile. Consider, for example, $\ell$-repellence: Given a game $G$ and a strategy profile $s_N$, find the maximum $c$ such that $s_N$ is $c$-repellent. We can define similar sub-problems for all security equilibria definitions.

\textbf{Computing Nash Equilibrium is PPAD-Complete.} \cite{ppad-more-than-2} showed that the problem of computing a Nash equilibrium with at least three players is PPAD-complete. \cite{ppad-2} showed that the problem is also PPAD-complete for two players.

The class of PPAD problems contains computational problems for which polynomial-time algorithms are not known to exist. Therefore, PPAD problems are considered intractable.

Some of our security equilibria, like $k$-resiliency and $\ell$-repellence, are also Nash Equilibria, so computing them is also PPAD-complete. Therefore, one should not expect our implementations to be efficient.

\textbf{$k$-resiliency and $\ell$-repellence.} Algorithm \ref{alg/k-l} finds the maximal $k$ (or $\ell$), in $O(2^n \times S \times n)$ time, where $n$ is the number of players, and $S$ is the total number of strategy profiles. The two outer for loops go over all possible coalitions (combinations) of $N$, which is $O(2^n)$. Then we loop over all possible strategy profiles for each coalition, which is $O(S)$. For $k$-resiliency, we loop over the members of the current coalition, which is $O(n)$. For $\ell$-repellence, we compute the coalition utility of each coalitions, which internally calls the utility functions of all the coalition members, which is also $O(n)$. Note that $S = \prod_{i} |A_i|$, so S is exponential in the number of players and number of actions.

\begin{algorithm}[htbp]
    \begin{algorithmic}[1]
        \STATE \textbf{INPUT:} $G$, $s_N$, $type$
        \STATE $N \leftarrow G.players$
        \STATE $n \leftarrow |N|$
        \FOR {$c \leftarrow 1, n$}
            \FOR {$C \in N.combinationsOfSize(c)$}
                \STATE $s_{-C} \leftarrow G.remainingStrategies(C)$
                \FOR {$s_C^\prime \in G.possibleStrategies(C)$}
                    \IF  {$type = resiliency$}
                        \FOR {$i \in C$}
                            \IF {$U_i(s_N) < U_i(s_C^\prime \cup s_{-C})$}
                                \STATE \textbf{return} $c - 1$
                            \ENDIF
                        \ENDFOR
                    \ENDIF
                    \IF  {$type = repellence$}
                        \IF {$U_C(s_N) < U_C(s_C^\prime \cup s_{-C})$}
                            \STATE \textbf{return} $c - 1$
                        \ENDIF
                    \ENDIF
                \ENDFOR
            \ENDFOR
        \ENDFOR
        \STATE \textbf{return} $n$
    \end{algorithmic}
    \caption{Given a game $G$ and a strategy profile $s_N$, find the maximum $c$ such that $s_N$ is $c$-resilient/repellent.}
    \label{alg/k-l}
\end{algorithm}

\textbf{$t$-immunity.} Algorithm \ref{alg/t-immunity} finds the maximal $t$ in $O(2^n \times S \times n)$ time. The two outer for loops go over all possible groups of Byzantine players (combinations) of $N$, which is $O(2^n)$. Then we loop over all possible strategy profiles for each group, which is $O(S)$. Finally, we loop over the players outside of the current Byzantine group, which is $O(n)$.

\begin{algorithm}[htbp]
    \begin{algorithmic}[1]
        \STATE \textbf{INPUT:} $G$, $s_N$
        \STATE $N \leftarrow G.players$
        \STATE $n \leftarrow |N|$
        \FOR {$c \leftarrow 1, n$}
            \FOR {$T \in N.combinationsOfSize(c)$}
                \STATE $s_{-T} \leftarrow G.remainingStrategies(T)$
                \FOR {$s_T^\prime \in G.possibleStrategies(T)$}
                    \FOR {$i \notin C$}
                        \IF {$U_i(s_T^\prime \cup s_{-T}) < U_i(s_N)$}
                            \STATE \textbf{return} $c - 1$
                        \ENDIF
                    \ENDFOR
                \ENDFOR
            \ENDFOR
        \ENDFOR
        \STATE \textbf{return} $n$
    \end{algorithmic}
    \caption{Given a game $G$ and a strategy profile $s_N$, find the maximum $c$ such that $s_N$ is $c$-immune.}
    \label{alg/t-immunity}
\end{algorithm}

\textbf{$(k, t)$-robustness and $(\ell, t)$-resistance.} Algorithm \ref{alg/k-l-t} finds maximal $k,t,\ell$ in $O(2^{2n} \times S^2 \times n)$ time. The two outer for loops go over all possible coalitions (combinations) of $N$, then we loop over all possible groups of Byzantine players (combinations) of $N - C$, which is $O(2^{2n})$. Then we loop over all possible strategy profiles for each coalition and each Byzantine group, which is $O(S^2)$. For $(k, t)$-robustness, we loop over the members of the current coalition, which is $O(n)$. For $(\ell, t)$-resistance, we compute the coalition utility of each coalitions, which is also $O(n)$.

\begin{algorithm}[htbp]
    \begin{algorithmic}[1]
        \STATE \textbf{INPUT:} $G$, $s_N$, $type$
        \STATE $S \leftarrow \{ \}$
        \STATE $N \leftarrow G.players$
        \STATE $n \leftarrow |N|$
        \STATE \textbf{LABEL:} Outer Loop
        \FOR {$c_1 \leftarrow 1, n - 1$}
            \FOR {$C \in N.combinationsOfSize(c_1)$}
                \FOR {$c_2 \leftarrow 1, n - c_1$}
                    \FOR {$T \in (N - C).combinationsOfSize(c_2)$}
                        \STATE $s_{-(C \cup T)} \leftarrow G.remainingStrategies(C \cup T)$
                        \FOR {$s_T^\prime \in G.possibleStrategies(T)$}
                            \FOR {$s_C^\prime \in G.possibleStrategies(C)$}
                                \IF  {$type = robustness$}
                                    \FOR {$i \in C$}
                                        \IF {$U_i(s_T^\prime \cup s_{-T}) < U_i(s_T^\prime \cup s_C^\prime \cup s_{-(C \cup T)})$}
                                            \IF {$t = 1$}
                                                \STATE \textbf{return} $S$
                                            \ENDIF
                                            \STATE $S.append((c_1, c_2 - 1))$
                                            \STATE \textbf{continue} Outer Loop
                                        \ENDIF
                                    \ENDFOR
                                \ENDIF
                                \IF  {$type = resistance$}
                                    \IF {$U_C(s_T^\prime \cup s_{-T}) < U_C(s_T^\prime \cup s_C^\prime \cup s_{-(C \cup T)})$}
                                        \IF {$t = 1$}
                                            \STATE \textbf{return} $S$
                                        \ENDIF
                                        \STATE $S.append((c_1, c_2 - 1))$
                                        \STATE \textbf{continue} Outer Loop
                                    \ENDIF
                                \ENDIF
                            \ENDFOR
                        \ENDFOR
                    \ENDFOR
                \ENDFOR
            \ENDFOR
            \STATE $S.append((c_1, n - c_1))$ 
        \ENDFOR
        \STATE \textbf{return} $S$
    \end{algorithmic}
    \caption{Given a game $G$ and a strategy profile $s_N$, find the set of maximal $c_1$ and $c_2$ such that $s_N$ is $(c_1, c_2)$-robust/resistant.}
    \label{alg/k-l-t}
\end{algorithm}

\textbf{$m$-stability.} 
Algorithm \ref{alg/m-stability} finds the maximal $m$ in $O(2^n \times S^2 \times n)$ time. The two outer for loops go over all possible coalitions (combinations) of $N$, which is $O(2^n)$. Then we loop over all possible strategy profiles for players outside of the coalition, which is $O(S)$. We then loop over all possible strategy profile for players of the coalition, which is $O(S)$ as well. Finally, we compute the coalition utility of each coalitions, which is $O(n)$.

\begin{algorithm}[htbp]
    \begin{algorithmic}[1]
        \STATE \textbf{INPUT:} $G$, $s_N$
        \STATE $N \leftarrow G.players$
        \STATE $n \leftarrow |N|$
        \FOR {$c \leftarrow 1, n$}
            \FOR {$C \in N.combinationsOfSize(c)$}
                \STATE $s_C \leftarrow G.remainingStrategies(N - C)$
                \FOR {$s_{-C}^\prime \in G.possibleStrategies(N - C)$}
                    \FOR {$s_C^\prime \in G.possibleStrategies(C)$}
                        \IF {$U_C(s_C \cup s_{-C}^\prime) < U_C(s_C^\prime \cup s_{-C}^\prime)$}
                            \STATE \textbf{return} $c - 1$
                        \ENDIF
                    \ENDFOR
                \ENDFOR
            \ENDFOR
        \ENDFOR
        \STATE \textbf{return} $n$
    \end{algorithmic}
    \caption{Given a game $G$ and a strategy profile $s_N$, find the maximum $c$ such that $s_N$ is $m$-stable.}
    \label{alg/m-stability}
\end{algorithm}


\section{Experimental Results}

\subsection{Multi-Party Games}

In this section, we briefly describe two games that we use in our experiments.

\textbf{Incentivized Outsourced Computation (IOC) \cite{ioc-game}:} The authors of \cite{ioc-game} defined an Incentivized Outsourced Computation (IOC) game. The game involves a boss and $n$ rational contractors as players. The boss wants to outsource the execution of a costly algorithm to the contractors. Each contractor has the option to choose between the diligent strategy and the lazy strategy. The diligent strategy involves running the correct algorithm, which costs $c(1)$, while the lazy strategy uses a less costly algorithm known as the "$q$ algorithm", which costs $c(q)$. The $q$ algorithm has a probability $q$ of producing the correct output and is assumed to be the same for all lazy players. If all contractors provide the same output, the boss accepts it as correct and rewards each contractor with reward $r$. However, if there is a difference in the outputs, the diligent contractors collaborate with the boss to identify and penalize the lazy ones. In this case, the diligent contractors receive the reward $r$ and an additional bounty $b$, while each contractor contributes a share towards the total bounties and incurs a fixed fine $f$. Table \ref{table/ioc-payoff} illustrates the resulting expected utility matrix. Note that for this mechanism to be meaningful, it is necessary that $c(q) < c(1) < r$.

\begin{table}[htbp]
    \caption{IOC Payoff Matrix (where $0 < k < n$)}
    \centering
    \resizebox{\linewidth}{!}{
        \begin{tabular}{|c|c|c|}
         \hline
         Others / This & Diligent & Lazy\\
         \hline
         All diligent & $r - c(1)$ & $rq - (f + b(n - 1))(1 - q) - c(q)$\\
         \hline
         $k$ lazy & $r + b(1 - q) - c(1)$ & $rq - (f + \frac{b(n - k - 1)}{k + 1})(1 - q) - c(q)$\\
         \hline
         All lazy & $r + b(1 - q) - c(1)$ & $r - c(q)$\\
         \hline
        \end{tabular}
    }
    \label{table/ioc-payoff}
\end{table}

The desired outcome of the game is all players choosing the diligent strategy. 
\cite{new-definitions} proved that IOC is not $2$-resilient. They also proved that for $n > 2$, if the boss sets the reward and bounty as $r (n - 1) / (n - 2) \geq b > r / (1 - q)$, then IOC is $(n - 1)$-stable. Moreover, they proved that IOC is $(n - 1)$-immune.

\textbf{Forwarding Dilemma (FD) \cite{fd-game}:} The authors of \cite{fd-game} introduced the Forwarding Dilemma (FD) game as a model to study the forwarding behavior of flooded packets in wireless ad hoc networks. Network nodes are the players of FD. Each node receives the same flooded packet. Each player can either forward the packet or drop it. Two factors determine the utility of each player: the network gain factor $g$ and the forwarding cost $c$. Table \ref{table/fd-payoff} shows the specific utilities for each player, considering their strategy and the strategies of others. It is essential to have $c < g$.

\begin{table}[htbp]
    \caption{FD Payoff Matrix}
    \centering
    \begin{tabular}{|c|c|c|}
     \hline
     Others / This & Forward & Drop\\
     \hline
     All drop & $g - c$ & $0$\\
     \hline
     At least one forward & $g - c$ & $g$\\
     \hline
    \end{tabular}
    \label{table/fd-payoff}
\end{table}

The desirable strategies in this game are when one player forwards the packet, and the rest drops it. 
\cite{new-definitions} proved that FD is not $2$-resilient, is $n$-repellent, is not $(1, 1)$-resistant, is not $(1, 1)$-robust, is not $1$-immune, and is not $1$-stable.

\subsection{Input Format}

We implemented \toolname{} in Python, using the Python interface to the Gambit\footnote{\url{https://github.com/gambitproject/gambit}} library. The source code is openly available on \href{https://github.com/CRYPTO-KU/GamuBlue-Game-Theory-Equilibrium-Finder}{GitHub}.

The Gambit library is a powerful tool for game-theory analysis, offering a range of features. One of its key strengths is its implementation of various algorithms for computing Nash Equilibria, providing researchers and practitioners with efficient methods to analyze strategic interactions. Additionally, Gambit includes a user-friendly graphical user interface (GUI) that facilitates the examination of small games, allowing users to visualize game structures and explore their strategic dynamics. Moreover, Gambit offers a Python interface, enabling developers to leverage its game representations and develop custom game-theory algorithms. This flexibility and accessibility make the Gambit library a valuable resource for studying and understanding game-theoretic concepts and applications.

\toolname{} accepts the NFG and AGG game representations defined by the Gambit library. The core difference is that AGG tends to be much more compact than NFG for highly structured games. The Gambit library's documentation contains a detailed explanation of both representations. Figures \ref{inp/ioc-nfg} and \ref{inp/ioc-agg} show example input representations for IOC with $N = 3$, $cost(1) = 10$, $cost(q) = 5$, $q = 0.5$, $r = 20$, $b = 20$, and $f = 2.5$ in NFG and AGG formats, respectively. Figures \ref{inp/fd-nfg} and \ref{inp/fd-agg} show example input representations for FD with $N = 3$, $g = 2$, and $c = 1$ in NFG and AGG formats, respectively.

\begin{figure}[htbp]
    \centering
    \verbatimboxed{inputs/ioc.nfg}
    \caption{IOC Represented in NFG Format For 3 Players}
    \label{inp/ioc-nfg}
\end{figure}

\begin{figure}[htbp]
    \centering
    \verbatimboxed{inputs/ioc.agg}
    \caption{IOC Represented in AGG Format For 3 Players}
    \label{inp/ioc-agg}
\end{figure}

\begin{figure}[htbp]
    \centering
    \verbatimboxed{inputs/fd.nfg}
    \caption{FD Represented in NFG Format For 3 Players}
    \label{inp/fd-nfg}
\end{figure}

\begin{figure}[htbp]
    \centering
    \verbatimboxed{inputs/fd.agg}
    \caption{FD Represented in AGG Format For 3 Players}
    \label{inp/fd-agg}
\end{figure}

\subsection{Timing Benchmarks}

\textbf{Experimental Setup.} We conducted the experiments on a machine with an Intel Core i7-8750H 2.20GHz processor, 16GB of DDR4 RAM, and 512GB of SSD storage. We ran each algorithm on each input 100 times. We report the mean running times of these runs.

\begin{figure}[htbp]
    \begin{tikzpicture}
        \begin{semilogyaxis}[
            xlabel = {Number of players $n$},
            ylabel = {Mean running time (sec)}
        ]
            \addlegendentry{IOC}
            \addplot table [x=n, y=mean, col sep=comma] {timings/ioc-k-resiliency.csv};
            \addlegendentry{FD}
            \addplot table [x=n, y=mean, col sep=comma] {timings/fd-k-resiliency.csv};
        \end{semilogyaxis}
    \end{tikzpicture}
    \caption{Experimental Performance of $k$-resiliency}
    \label{perf/k-reiliency}
\end{figure}
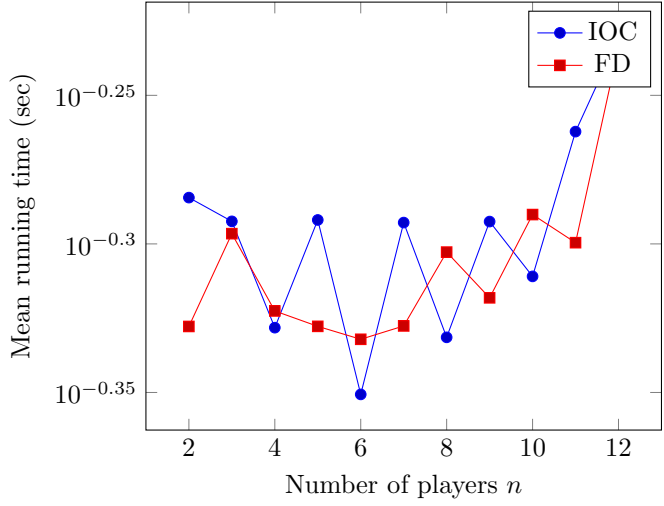

Figure \ref{perf/k-reiliency} for $k$-resiliency shows that Algorithm \ref{alg/k-l} terminates quickly (less than one second for $12$ players). These timings are due to neither IOC nor FD being 2-resilient, as shown by \cite{new-definitions}.  The timings have fluctuations because external operating system factors have a significant impact on this small scale.

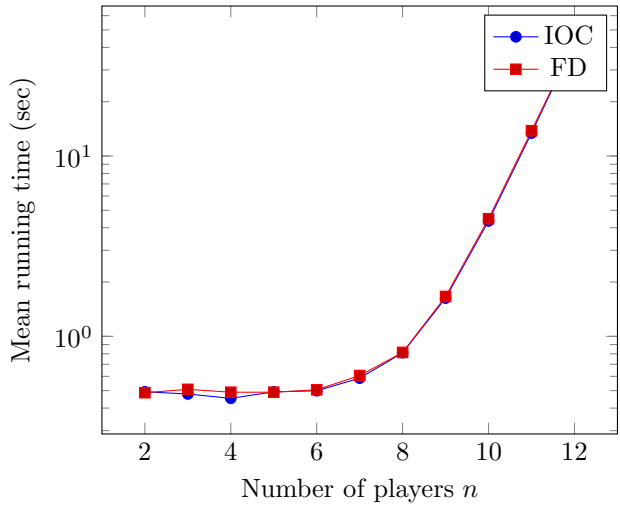
\begin{figure}[htbp]
    \begin{tikzpicture}
        \begin{semilogyaxis}[
            xlabel = {Number of players $n$},
            ylabel = {Mean running time (sec)}
        ]
            \addlegendentry{IOC}
            \addplot table [x=n, y=mean, col sep=comma] {timings/ioc-l-repellence.csv};
            \addlegendentry{FD}
            \addplot table [x=n, y=mean, col sep=comma] {timings/fd-l-repellence.csv};
        \end{semilogyaxis}
    \end{tikzpicture}
    \caption{Experimental Performance of $\ell$-repellence}
    \label{perf/l-repellence}
\end{figure}

\cite{new-definitions} proved that IOC is $(n - 1)$-repellent and FD is $n$-repellent. The timings in Figure \ref{perf/l-repellence} align with this, and they also show the exponential nature of 
Algorithm \ref{alg/k-l}.

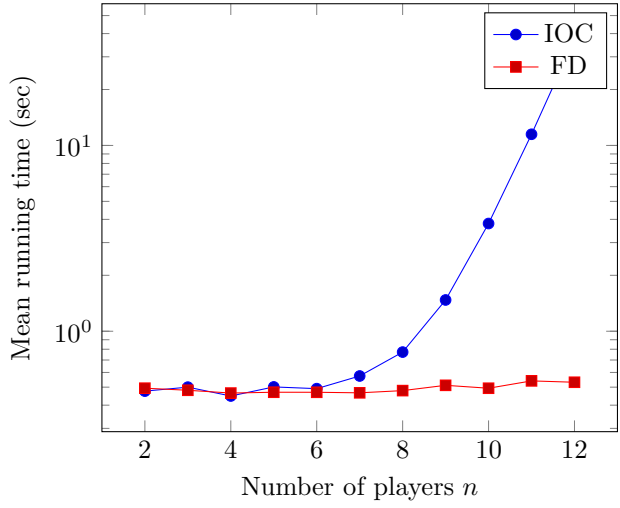
\begin{figure}[htbp]
    \begin{tikzpicture}
        \begin{semilogyaxis}[
            xlabel = {Number of players $n$},
            ylabel = {Mean running time (sec)}
        ]
            \addlegendentry{IOC}
            \addplot table [x=n, y=mean, col sep=comma] {timings/ioc-t-immunity.csv};
            \addlegendentry{FD}
            \addplot table [x=n, y=mean, col sep=comma] {timings/fd-t-immunity.csv};
        \end{semilogyaxis}
    \end{tikzpicture}
    \caption{Experimental Performance of $t$-immunity}
    \label{perf/t-immunity}
\end{figure}

In Figure \ref{perf/t-immunity}, we can see the rapid growth of Algorithm \ref{alg/t-immunity} timings in IOC as the number of players increases. However, Algorithm \ref{alg/t-immunity} does not seem to get slower on FD. The reason is that IOC is $(n - 1)$-immune, while FD is not even $1$-immune, as proven by \cite{new-definitions}, and hence the algorithm terminates early finding this fact.

\begin{figure}[htbp]
    \begin{tikzpicture}
        \begin{semilogyaxis}[
            xlabel = {Number of players $n$},
            ylabel = {Mean running time (sec)}
        ]
            \addlegendentry{IOC}
            \addplot table [x=n, y=mean, col sep=comma] {timings/ioc-k-t-robustness.csv};
            \addlegendentry{FD}
            \addplot table [x=n, y=mean, col sep=comma] {timings/fd-k-t-robustness.csv};
        \end{semilogyaxis}
    \end{tikzpicture}
    \caption{Experimental Performance of $(k, t)$-robustness}
    \label{perf/k-t-robustness}
\end{figure}
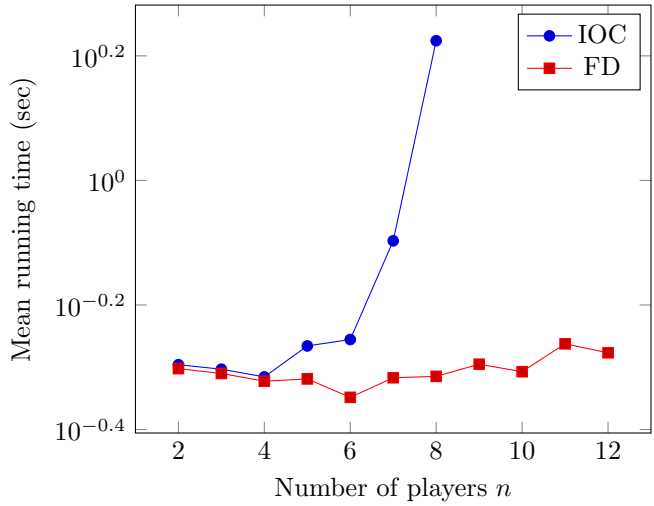

Similarly in Figure \ref{perf/k-t-robustness}, we can see the rapid growth of Algorithm \ref{alg/k-l-t} timings in IOC as the number of players increases. However, Algorithm \ref{alg/k-l-t} does not seem to get slower on FD. The reason is that FD is not $(1, 1)$-robust, as proven by \cite{new-definitions}, while the experiments show that IOC is $(1, n - 1)$-robust.

\begin{figure}[htbp]
    \begin{tikzpicture}
        \begin{semilogyaxis}[
            xlabel = {Number of players $n$},
            ylabel = {Mean running time (sec)}
        ]
            \addlegendentry{IOC}
            \addplot table [x=n, y=mean, col sep=comma] {timings/ioc-l-t-resistance.csv};
            \addlegendentry{FD}
            \addplot table [x=n, y=mean, col sep=comma] {timings/fd-l-t-resistance.csv};
        \end{semilogyaxis}
    \end{tikzpicture}
    \caption{Experimental Performance of $(\ell, t)$-resistance}
    \label{perf/l-t-resistance}
\end{figure}
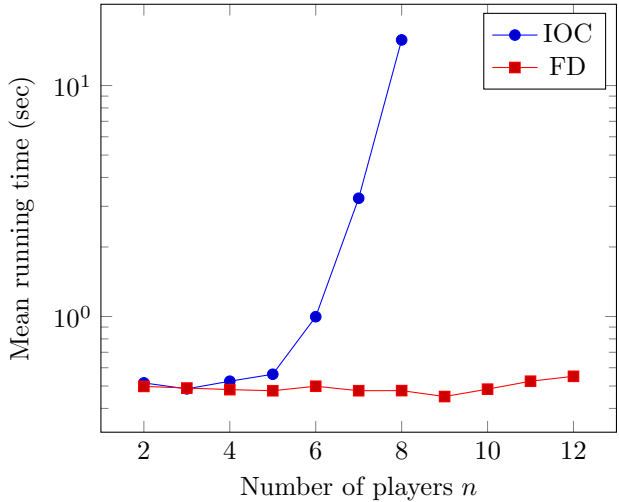

Again in Figure \ref{perf/l-t-resistance}, we can see the rapid growth of Algorithm \ref{alg/k-l-t} timings in IOC as the number of players increases. However, Algorithm \ref{alg/k-l-t} does not seem to get slower on FD. The reason is that IOC is $(\ell, n - \ell)$-resistant for all $\ell \leq n - 1$, while FD is not even $(1, 1)$-resistant, as proven by \cite{new-definitions}.

\begin{figure}[htbp]
    \begin{tikzpicture}
        \begin{semilogyaxis}[
            xlabel = {Number of players $n$},
            ylabel = {Mean running time (sec)}
        ]
            \addlegendentry{IOC}
            \addplot table [x=n, y=mean, col sep=comma] {timings/ioc-m-stability.csv};
            \addlegendentry{FD}
            \addplot table [x=n, y=mean, col sep=comma] {timings/fd-m-stability.csv};
        \end{semilogyaxis}
    \end{tikzpicture}
    \caption{Experimental Performance of $m$-stability}
    \label{perf/m-stability}
\end{figure}
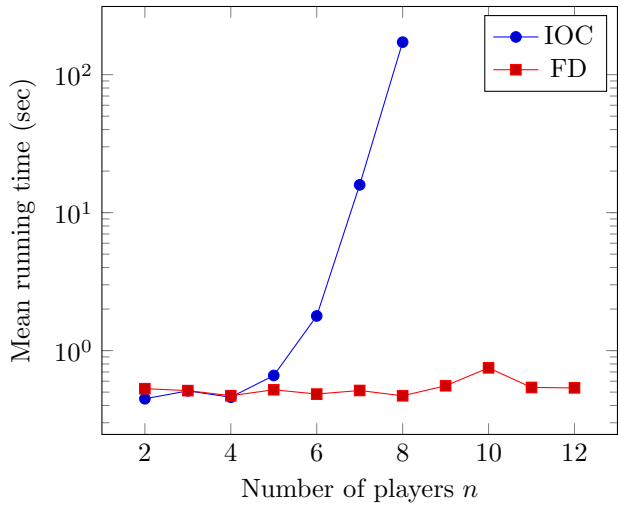

In Figure \ref{perf/m-stability}, we can see the rapid growth of Algorithm \ref{alg/m-stability} timings in IOC as the number of players increases. However, Algorithm \ref{alg/m-stability} does not seem to get slower on FD. The reason is that IOC is $(n - 1)$-stable, while FD is not even $1$-stable, as proven by \cite{new-definitions}. All these results confirm that our algorithms return the maximal $k,\ell,t,m$ values as soon as they are found.

\section{Conclusion}

We implemented a tool for computing equilibria definitions: $k$-resiliency, $t$-immunity, $(k, t)$-robustness, $\ell$-repellence, $(\ell, t)$-resistance, and $m$-stability. We also analyzed the worst-case time complexity of each of the algorithms. We conducted experiments on the Incentivized Outsourced Computation (IOC) and Forwarding Dilemma (FD) games. Our tool is available open source:  \href{https://github.com/CRYPTO-KU/GamuBlue-Game-Theory-Equilibrium-Finder}{github.com/CRYPTO-KU/GamuBlue-Game-Theory-Equilibrium-Finder}

\section*{Acknowledgements}
We thank TÜBİTAK (the Scientific and Technological Research Council of Turkey) project 119E088.


\bibliographystyle{IEEEtran}
\bibliography{IEEEabrv, references}

\end{document}